\documentclass[aps,preprint,nofootinbib,superscriptaddress,prc]{revtex4}
\usepackage{amssymb}

\usepackage{epsfig}
\usepackage[bookmarksnumbered,bookmarksopen,colorlinks,citecolor=blue,linkcolor=blue] {hyperref}
\begin{document}


\title{Systematics of fusion probability in  ``hot" fusion reactions}

\author{Ning Wang}
\affiliation{ Department of Physics, Guangxi Normal University,
Guilin 541004, People's Republic of China }

\author{Junlong Tian}
 \email{tianjunlong@gmail.com}
 \affiliation{ School of Physics and Electrical Engineering,
Anyang Normal University, Anyang 455002, People's Republic of
China }

\author{Werner Scheid}
\affiliation{Institut f\"{u}r Theoretische Physik der
Universit\"{a}t, D-35392 Giessen, Germany}

\begin{abstract}
The fusion probability in ``hot" fusion reactions leading to the
synthesis of super-heavy nuclei is investigated systematically.
The quasi-fission barrier influences the formation of the
super-heavy nucleus around the ``island of stability" in addition
to the shell correction. Based on the quasi-fission barrier height
obtained with the Skyrme energy-density functional, we propose an
analytical expression for the description of the fusion
probability, with which the measured evaporation residual cross
sections can be reproduced acceptably well. Simultaneously, some
special fusion reactions for synthesizing new elements 119 and 120
are studied. The predicted evaporation residual cross sections for
$^{50}$Ti+$^{249}$Bk are $\sim 10-150$ fb at energies around the
entrance-channel Coulomb barrier. For the fusion reactions
synthesizing element 120 with projectiles $^{54}$Cr and $^{58}$Fe,
the cross sections fall to a few femtobarns which seems beyond the
limit of the available facilities.

\end{abstract}

\maketitle

Synthesis of super-heavy nuclei (SHN) through fusion reactions is
a field of very intense studies in the recent decades
\cite{Hof00,Ogan00,Ogan04,Mori04a,Ogan06,Ogan10,Naza,Sob,shen11,bao,Gup05,Zag08,Ada09,Siw07,Nas,Huang,Zhou11}.
Up to now, the superheavy elements $Z=107 \sim 118$ have already
been synthesized \cite{Hof00,Ogan00,Ogan04,Mori04a,Ogan06,Ogan10}
through ``cold" fusion reactions that use lead or bismuth targets
with appropriate projectiles following the emission of one or two
neutrons from a ``cold" compound system, or ``hot" fusion
reactions that use actinide targets from uranium to californium
with beams of $^{48}$Ca following the evaporation of 3 to 5
neutrons from a ``hot" system. The reaction $^{50}$Ti+$^{249}$Cf
for producing element 120 is currently being studied in GSI
without a result up to now \cite{Dull}. Besides the experiments,
both the structure of SHN and the mechanism of fusion reactions
are also being intensively investigated theoretically. On one
hand, the precise calculation of the masses and the shell
correction of SHN which plays a key role in determining the
fission barrier and the center of the ``island of stability" for
SHN, is of great importance. Recently, Liu et al. \cite{Wang11}
proposed an improved macroscopic-microscopic mass model (also
called Weizs\"acker-Skyrme mass model) with an rms error of 336
keV with respect to the 2149 known masses and 248 keV to the
measured $\alpha$-decay energies of 46 super-heavy nuclei, with
which they found that the SHN with the largest shell corrections
are located around $Z= 116\sim120$ and $N=178$ rather than at
$N=184$. It is found that the difference of the calculated
evaporation residual cross sections for reactions leading to
element 120 reaches two orders of magnitude by adopting two
different mass tables for SHN \cite{Nas}. It is therefore
interesting to investigate the production cross sections of SHN in
fusion reactions by using the Weizs\"acker-Skyrme mass model.

On the other hand, the fusion probability of heavy nuclei and
super-heavy nuclei should be systematically investigated for
testing the models and for giving reliable predictions of fusion
reactions leading to new super-heavy elements. Theoretical support
for these very time-consuming and extremely-expensive experiments
is vital for choosing the optimum target-projectile-energy
combinations and for the estimation of cross sections. In the
practical calculation of the evaporation residue cross section,
the reaction process leading to the synthesis of SHN can be
divided into three steps. Firstly, the projectile is captured by
the target and a dinuclear system is formed which then evolves
into the compound nucleus, and finally, the compound nucleus loses
its excitation energy mainly by the emission of particles and
$\gamma$-rays and goes to its ground state. The simplified version
of the evaporation residue cross section is given by
\cite{Wang08,Siw07}
\begin{eqnarray}
\sigma_{\rm ER}(E_{\rm c.m.}) = \sigma_{\rm cap}(E_{\rm c.m.})
P_{\rm CN}(E_{\rm c.m.}) W_{\rm sur}(E_{\rm c.m.}).
\end{eqnarray}
Here, $\sigma_{\rm cap}$, $P_{\rm CN}$ and $W_{\rm sur}$ denote
the capture cross section of the colliding nuclei overcoming the
Coulomb barrier, the probability of the compound nucleus formation
(i.e., the fusion probability) after the capture and the survival
probability of the excited compound nucleus, respectively. The
most unclear part may be the value of $P_{\rm CN}$ which is
usually calculated by different models for the dynamics
\cite{Ada09,bao,Zag08} based on the potential energy surface of
the reaction system or by empirical formulas \cite{Love,Zag08}. In
Refs.\cite{Liu06,Wang06,Wang08,Wang09}, the methods for the
calculation of $\sigma_{\rm cap}$ and $W_{\rm sur}$ are well
established in general, with the Skyrme energy-density functional
and an empirical barrier distribution for describing the capture
cross sections and with the HIVAP code \cite{Reis81,Reis85,Reis92}
for describing the survival probability of the compound nuclei.
Based on the previous work \cite{Wang08}, the fusion probability
in reactions leading to SHN will be further systematically
investigated in this work by applying the Weizs\"acker-Skyrme mass
model for describing the masses of SHN.

\begin{figure}
\includegraphics[angle=-0,width= 0.65\textwidth]{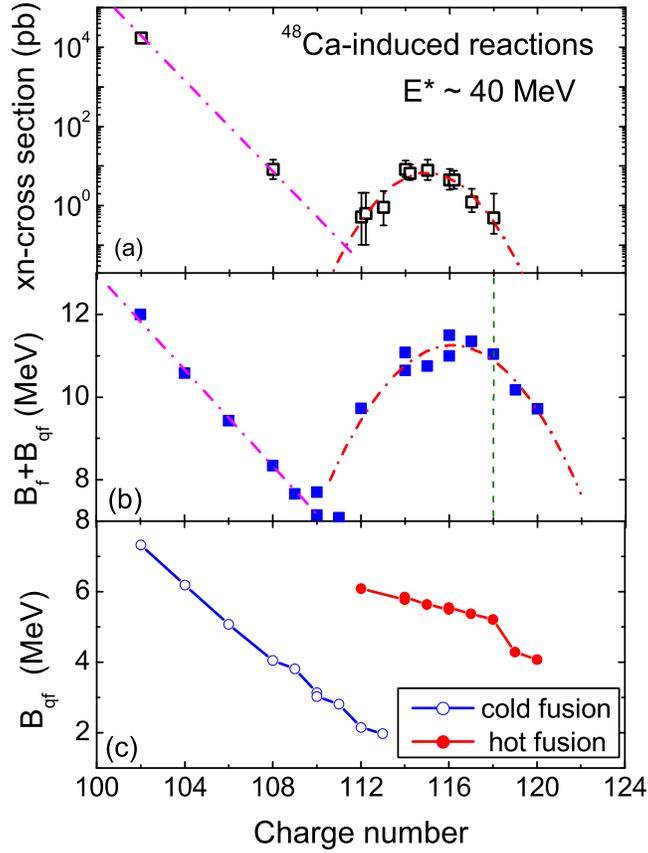}
 \caption{(Color online) (a) Measured evaporation residual
 cross-sections of some $^{48}$Ca-induced reactions. The data are extracted
 from a figure in Oganessian's talk \cite{Ogantalk}. (b) Sum of fission barrier height and quasi-fission
 barrier height for some fusion reactions leading to the synthesis of super-heavy nuclei as a function of the charge number of SHN.
 The dot-dashed curves are to guide the eyes. The vertical dashed line shows the position of $Z=118$
 and the squares on its right side denote the results for $^{50}$Ti+$^{249}$Bk and $^{50}$Ti+$^{249}$Cf. (3) Quasi-fission barrier height for these reactions.
 The open and filled circles denote the results for ``cold" and ``hot" fusion reactions, respectively.  }
\end{figure}

We first investigate the relation between the fission barrier
height, the quasi-fission barrier height and the evaporation
residual cross sections for some fusion reactions leading to SHN.
The fission barrier height $B_{\rm f}$ of a nucleus can be
approximately estimated by the shell correction $\Delta E^{\rm
shell}$ and the macroscopic fission barrier $B_{\rm f}^{\rm Mac}$
of the nucleus \cite{Wang08}, i.e.,
\begin{eqnarray}
 B_{\rm f} \approx B_{\rm f}^{\rm Mac}+\Delta E^{\rm Shell}=B_{\rm f}^{\rm Mac}+(E_{\rm exp}-E_{\rm
 LD}).
\end{eqnarray}
Here, $E_{\rm exp}$ and $E_{\rm LD}$ denote the measured binding
energy and the liquid drop energy of the nucleus (positive
values), respectively. Fig. 1(a) shows the measured evaporation
residual cross sections for some $^{48}$Ca-induced reactions as a
function of the charge number of the compound
 nucleus. The data are extracted from a figure in Oganessian's talk
 \cite{Ogantalk}, and these data are also shown in Ref.\cite{Pop}.
 At $Z\approx114-116$ there exists a peak for the cross sections, which seems to be an evidence of the center of the
``island of stability" for super-heavy nuclei being located at
$Z\leqslant  120$.
 Fig. 1(b) shows the sum of the fission barrier height and the quasi-fission
 barrier height for some fusion reactions leading to the synthesis of SHN.
 Here, the quasi-fission barrier height $B_{\rm qf}$ is defined as the depth of the capture pocket
 in the entrance-channel potential (see the sub-figure in Fig. 2), which is calculated by using
 the Skyrme energy-density functional \cite{Bart82} together with
 the extended Thomas-Fermi approximation including all terms up to the second order in the spatial
derivatives (ETF2) as mentioned in \cite{Liu06}. For
 calculating the shell corrections in Fig. 1(b), the obtained binding
 energies and the corresponding liquid-drop energies $E_{\rm LD}$ of the
 SHN in Ref.\cite{Wang11} are adopted.
One sees that the dependence of $B_{\rm f}+B_{\rm qf}$ on the
charge number is very similar to the dependence of the evaporation
residual cross sections on the charge number. The peak around
$Z\approx116$ in Fig.1(b) comes from the contributions of the
shell correction  and the quasi-fission barrier height. The
quasi-fission barrier height  decreases linearly with the charge
number of the compound nucleus  [see Fig. 1(c)], and the largest
shell corrections are located around $Z= 116\sim120$ according to
the Weizs\"acker-Skyrme mass model. The sum of the shell
correction $\Delta E^{\rm Shell}$ and the quasi-fission barrier
height results in the peak at $Z\approx114-116$ of the evaporation
residual cross sections. Fig. 1 indicates that the quasi-fission
barrier influences the formation of the super-heavy nuclei around
the known ``island of stability" in addition to the shell
correction. For very asymmetric fusion reactions leading to
intermediate and heavy nuclei rather than the SHN, the
quasi-fission barrier height is relatively high. With the increase
of the charge number of the projectile or target nuclei in the
fusion reactions leading to the synthesis of SHN, the
quasi-fission barrier height decreases gradually.

\begin{figure}
\includegraphics[angle=-0,width= 0.8\textwidth]{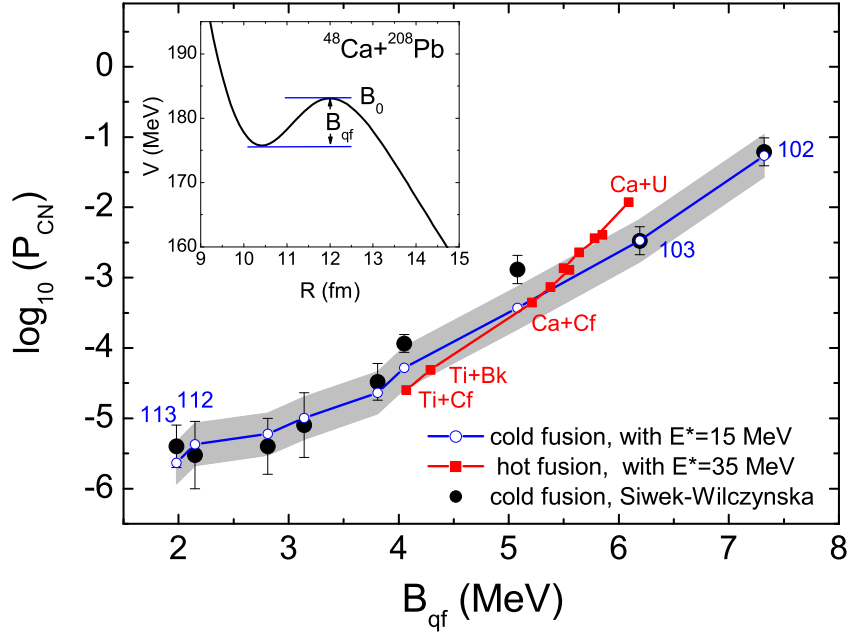}
 \caption{(Color online) Fusion probability in reactions leading to super-heavy nuclei as a function of quasi-fission barrier height.
 The open circles and the filled circles denote the
calculated results for ``cold" fusion reactions with excitation
energy $E^{*} =15$ MeV and the extracted fusion probability in
Ref.\cite{Siw07} for the same reactions, respectively. The solid
squares denote the results for ``hot" fusion reactions with $E^{*}
=35$ MeV. Sub-figure: Calculated entrance-channel potential for
$^{48}$Ca+$^{208}$Pb.
 $B_{\rm qf}$ and $B_0$ denote the depth of the pocket and the height of the Coulomb barrier, respectively.}
\end{figure}

It is known that the fission barrier height strongly influences
the survival probability of super-heavy nuclei. In addition, one
expects that the fusion probability increases with increasing of
the quasi-fission barrier height and the incident energies in the
reactions leading to the SHN \cite{Huang}. To consider the
contribution of the quasi-fission barrier, we propose an
analytical formula for description of the
 fusion probability $P_{\rm CN}$,
\begin{eqnarray}
P_{\rm CN}(E^{*})=\frac{1}{C}\exp \left( 3 B_{\rm qf}+0.3E^{*}
\right ).
\end{eqnarray}
Where, $E^{*}=E_{\rm c.m.}+Q$ denotes the excitation energy of the
compound nucleus (in MeV). In addition, we introduce a truncation
for the value of $P_{\rm CN}$, i.e. $P_{\rm CN}$ should not be
larger than 1.  The parameters in Eq. (3) with $C=\exp(50 |\eta|)$
are determined by fitting the measured evaporation residual cross
sections of $^{48}$Ca+$^{248}$Cm and $^{48}$Ca+$^{249}$Cf. Here,
$\eta=(A_1-A_2)/(A_1+A_2)$ denotes the mass asymmetry of the
reaction system. For $^{48}$Ca-induced ``hot" fusion reactions
leading to SHN, $|\eta| \approx 0.67$. The calculation for the
capture cross sections and the survival probability of compound
nucleus are the same as those in Ref.\cite{Wang08} except that the
masses of unknown nuclei are given by the Weizs\"acker-Skyrme mass
model \cite{Wang11} rather than by the finite range droplet model
(FRDM) \cite{Moll95}. Fig. 2 shows the fusion probability in
reactions leading to SHN as a function of the quasi-fission
barrier height $B _{\rm qf}$. The solid squares denote the results
for ``hot" fusion reactions with an excitation energy of $E^{*}
=35$ MeV. The open and filled circles denote the calculated
results for ``cold" fusion reactions with $E^{*} =15$ MeV and the
extracted fusion probability in Ref.\cite{Siw07} for the same
reactions, respectively. Here the parameter $C$ is slightly
different for the ``cold" fusion reactions which will be discussed
later.  The corresponding evaporation residual cross sections for
these reactions will be shown in Fig. 3, Fig. 4 and Fig. 5. The
shades show the estimated uncertainty of the calculated fusion
probability with Eq.(3). With decreasing of the quasi-fission
barrier height, the fusion probability decreases exponentially.
The obtained fusion probabilities in this work and those in
Ref.\cite{Siw07} are very close to each other for the ``cold"
fusion reactions. In addition, it seems that the obtained fusion
probability for the ``cold" fusion reaction and that for the
``hot" fusion reaction are comparable when their quasi-fission
barrier heights are close to each other.

\begin{figure}
\includegraphics[angle=-0,width= 0.8\textwidth]{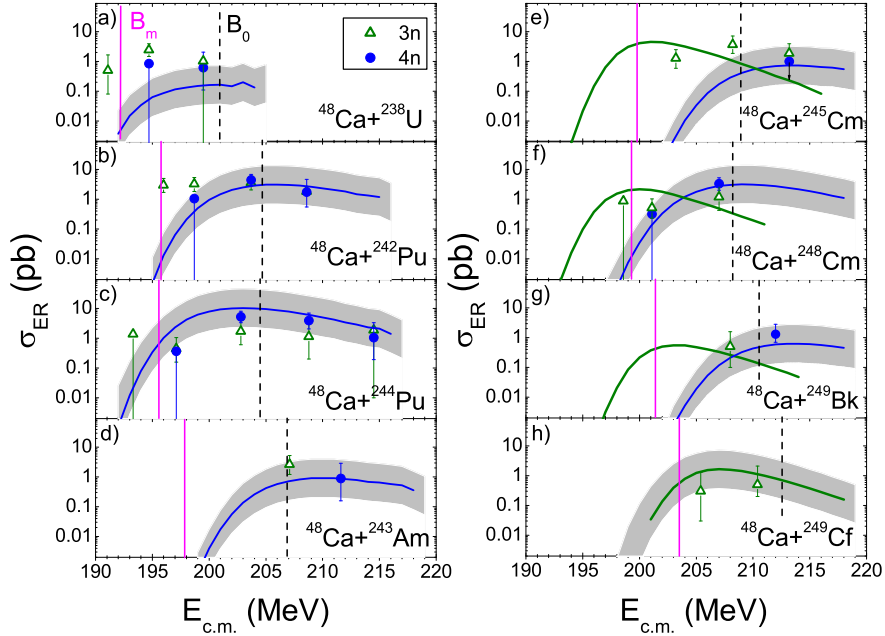}
 \caption{(Color online)  Evaporation residual cross sections for some ``hot" fusion
 reactions. The data are taken from Refs.\cite{Tian07,Ogan10}. The dashed and
 solid lines denote the entrance-channel Coulomb barrier height
$B_0$ under the sudden approximation for the densities and the
mean barrier height $B_{\rm m}$, respectively. The green and blue
curves denote the corresponding calculation results for the 3n and
4n channels with $C=\exp(50 |\eta|)$, respectively. The shades
show the uncertainty of the calculated $\sigma_{\rm ER}$ for the
4n channel [except (h) in which the shades are for the 3n
channel].}
\end{figure}

\begin{figure}
\includegraphics[angle=-0,width= 0.75\textwidth]{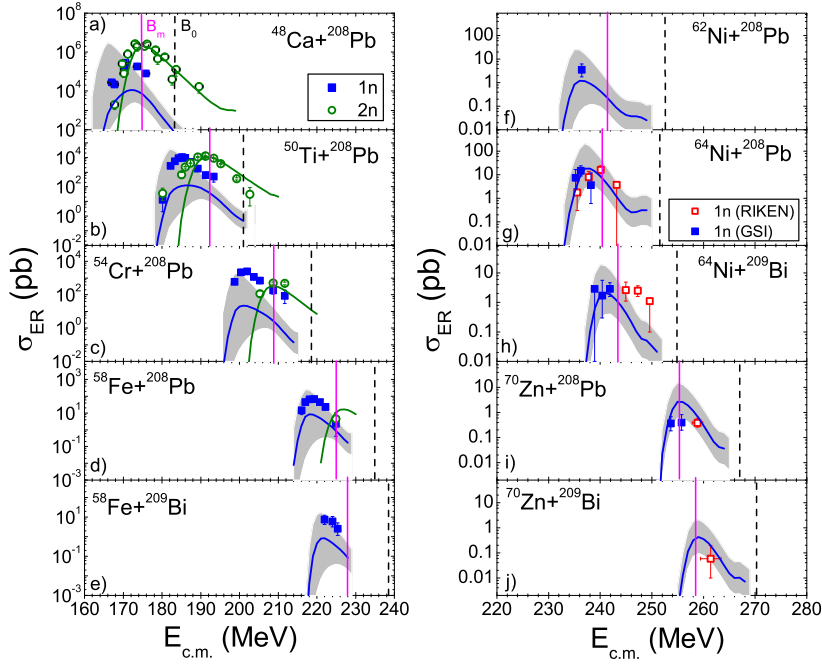}
 \caption{(Color online) The same as Fig.3, but for some ``cold" fusion reactions. Here, we adopt $C=\exp(47 |\eta|)$ in the calculations. The data are taken from Ref.\cite{Tian07}.
 The shades show
the uncertainty of the calculated $\sigma_{\rm ER}$ for the 1n
channel.}
\end{figure}

Fig. 3 shows the evaporation residual cross sections for some
``hot" fusion reactions. The open triangles and solid circles
denote the experimental data for the 3n and 4n channels,
respectively. The green and blue curves denote the corresponding
calculated results. The dashed and solid lines denote the
entrance-channel Coulomb barrier height $B_0$ from the Skyrme
energy-density functional and the mean barrier height $B_{\rm
m}\approx 0.956 B_0$ according to the empirical barrier
distribution \cite{Liu06}, respectively.  The shades show the
corresponding uncertainty of this approach due to the systematic
errors in the calculation of $\sigma_{\rm cap}$, $W_{\rm sur}$ and
$P_{\rm CN}$, which is about a factor of 4.4 for $\sigma_{\rm ER}$
at energies above the mean barrier height $B_{\rm m}$. The
estimated uncertainties \cite{Wang08} for $\sigma_{\rm cap}$,
$W_{\rm sur}$ and $P_{\rm CN}$ are about factors of 1.18, 1.85 and
2.0 at $E_{\rm c.m.}>B_{\rm m}$, respectively. We would like to
state that the parameters in Eq.(3) are determined just by
reactions $^{48}$Ca+$^{248}$Cm and $^{48}$Ca+$^{249}$Cf. With the
same parameters we find that the experimental data of other ``hot"
fusion reactions can also be reproduced reasonably well.

To further test the model, the ``cold" fusion reactions leading to
SHN are also investigated systematically. In Fig. 4, we show the
evaporation residual cross sections for some ``cold" fusion
reactions. The solid squares and open circles denote the
experimental data for the 1n and 2n channels, respectively. The
blue and green solid curves denote the corresponding calculated
results. Here, we slightly change the parameter $C=\exp(47
|\eta|)$ considering the differences between the ``cold " fusion
reactions and the ``hot" fusion reactions. Carrying out the same
calculations as for the``hot" fusion reactions, we can roughly
reproduce the measured evaporation residual cross sections for the
``cold" fusion reactions. The larger uncertainty for the ``cold"
fusion reactions at sub-barrier energies is due to the factor $g$
in the barrier distribution function for calculating the capture
cross sections \cite{Wang08}.

\begin{figure}
\includegraphics[angle=-0,width= 0.75\textwidth]{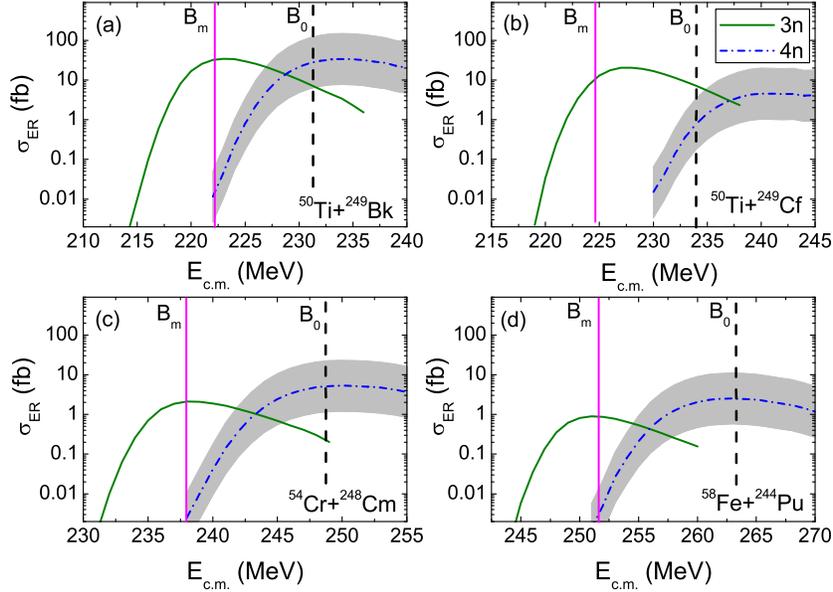}
 \caption{(Color online)  The same as Fig.3 but for fusion reactions
 leading to the synthesis of elements 119 and 120 (in femtobarn). The green solid and blue dot-dashed curves
 denote the predicted results for the 3n and 4n channels with $C=\exp(50 |\eta|)$, respectively.  }
\end{figure}
\begin{table}
\caption{ The predicted optimal evaporation residual cross section
(in femtobarn) for four reactions leading to elements 119 and
120.}
\begin{tabular}{ccccc}
\hline\hline
  Reference & ~~$^{50}$Ti+$^{249}$Bk~~  & ~~$^{50}$Ti+$^{249}$Cf ~~ & ~~$^{54}$Cr+$^{248}$Cm ~~ & ~~ $^{58}$Fe+$^{244}$Pu ~~  \\ \hline
Liu  \cite{bao}    & $\sim  600 $   & $\sim 100 $ & $-$ & $-$   \\
Nan Wang \cite{WN} &  $\sim 1000$  &  $\sim  200$    & $\sim 40 $
& $\sim 30 $ \\
Zagrebaev \cite{Zag08}&  $\sim  50 $  &  $\sim  40 $    & $\sim  20 $    & $\sim 5 $ \\
This work &  $\sim  35 $  &  $\sim  20 $    & $\sim  5 $ & $\sim
3 $ \\
 \hline\hline
\end{tabular}
\end{table}

The above calculations give us great confidence for investigating
the evaporation residual cross sections of fusion reactions
leading to the synthesis of new elements. Fig. 5 shows the
predicted evaporation residual cross sections for the reactions
$^{50}_{22}$Ti+$^{249}_{ \;\; 97}$Bk, $^{50}_{22}$Ti+$^{249}_{\;\;
98}$Cf, $^{54}_{24}$Cr+$^{248}_{ \;\; 96}$Cm and $^{58}_{
26}$Fe+$^{244}_{ \;\; 94}$Pu leading to the synthesis of the
elements 119 and 120. The solid and dashed curves denote the
results for the 3n and 4n channels, respectively.  For the
reaction $^{50} {\rm Ti}+ ^{249}{\rm Bk}\rightarrow ^{\; \;
295}119+4n$, the obtained evaporation residual cross sections
$\sigma_{\rm ER}$ are about $10 \sim 150 $ femtobarn at incident
energies around the entrance-channel Coulomb barrier $B_0$. For
the reactions synthesizing element 120, the calculated
$\sigma_{\rm ER}$ are smaller than those for element 119, and the
$\sigma_{\rm ER}$ falls to a few femtobarns for the latter two
reactions $^{54}$Cr+$^{248}$Cm and $^{58}$Fe+$^{244}$Pu. Here, we
would like to state that the parameter $C$ in Eq.(3) for the
description of the fusion probability depends on the parameter
sets for calculating the survival probability of super-heavy
nuclei. If one takes a different liquid-drop formula for
calculating the fission barrier, the parameter $C$ must be
re-adjusted. Adopting different formulas for describing the
liquid-drop energies \cite{Wang11,Myer67} and the binding energies
of yet unmeasured nuclei \cite{Wang11,Moll95} (the parameter $C$
is re-adjusted to fit the measured $\sigma_{\rm ER}$ of ``hot"
fusion reactions), we find that the calculated evaporation
residual cross sections for the four reactions leading to the
elements 119 and 120 are close to the corresponding results in
Fig. 5, with deviations which are smaller than the uncertainty of
this approach. To compare with the results from other models, we
list in Table I the calculated optimal evaporation residual cross
sections from four different models for the four reactions
mentioned in Fig. 5. One sees that for $^{50}$Ti+$^{249}$Bk and
$^{50}$Ti+$^{249}$Cf the results from Ref.\cite{bao} with the
fusion-by-diffusion model and Ref.\cite{WN} with the dinuclear
system model are in the same order of magnitude (hundreds of fb),
while the results from Ref.\cite{Zag08} and those in this work are
in the same order of magnitude (tens of fb). All models in Table I
predict that the optimal evaporation residual cross sections for
$^{50}{\rm Ti}+^{249}{\rm Cf}\rightarrow ^{\; \; 299-x{\rm n}}120$
are smaller than those for $^{50}{\rm Ti}+^{249}{\rm
Bk}\rightarrow ^{\; \; 299-x{\rm n}}119$.

In summary, the entrance-channel Coulomb barrier determines the
capture cross sections, the quasi-fission barrier due to the
pocket in the entrance-channel potential influences the fusion
probability and the fission barrier is related to the survival
probability, all play an important role for a reliable calculation
of the evaporation residual cross sections in fusion reactions
leading to the synthesis of SHN. Of course the dynamics of the
fusion process and the zero-point motion for fission process
\cite{Mosel,Smol} also influence the cross sections significantly,
but it is beyond the scope of this work. Based on the previously
established methods for the calculation of capture cross sections
and the survival probability, we further proposed an analytical
formula with only three fixed parameters for a systematic
description of the fusion probability in reactions leading to SHN.
With this formula the measured evaporation residual cross sections
for the ``hot" fusion reactions can be reproduced reasonably well.
The predicted evaporation residual cross sections $\sigma_{\rm
ER}$ for the reaction $^{50}$Ti+$^{249}$Bk are about $10\sim 150$
femtobarn at incident energies around the entrance-channel Coulomb
barrier, and the predicted $\sigma_{\rm ER}$ for the reactions
$^{54}$Cr+$^{248}$Cm and $^{58}$Fe+$^{244}$Pu falls to a few
femtobarns which seems beyond the limit of the available
facilities.

\begin{center}
\textbf{ACKNOWLEDGEMENTS}
\end{center}

We thank Shan-Gui Zhou for a careful reading of the manuscript,
and Jing-Dong Bao for a useful discussion. This work was supported
by National Natural Science Foundation of China, Nos 10875031,
10847004, 11005003 and 10979024.


\begin{thebibliography}{99}



\bibitem{Hof00} S. Hofmann and G. M\"unzenberg, Rev. Mod. Phys. \textbf{72}, 733
(2000).

\bibitem{Ogan00} Yu. Ts. Oganessian, V. K. Utyonkov, et al., Phys. Rev. C \textbf{62}, 041604(R) (2000).


\bibitem{Ogan04} Yu. Ts. Oganessian, V. K. Utyonkov, et al., Phys. Rev. C \textbf{69}, 021601(R) (2004).


\bibitem{Mori04a}  K. Morita, K. Morimoto, et al., J. Phys. Soci. Japan
\textbf{73}, 2593 (2004).

\bibitem{Ogan06} Yu. Ts. Oganessian, V. K. Utyonkov, et al., Phys. Rev. C \textbf{74}, 044602 (2006).

\bibitem{Ogan10} Yu. Ts. Oganessian et al.,  Phys. Rev. Lett.
\textbf{104}, 142502 (2010).

\bibitem{Naza} S. Cwiok, P. H. Heenen and W. Nazarewicz, Nature \textbf{433}, 705 (2005).

\bibitem{Sob} A. Sobiczewski,  K. Pomorski, Prog. Part. Nucl. Phys. \textbf{58},
292 (2007).

\bibitem{shen11} C. Shen, D. Boilley, et al., Phys. Rev. C \textbf{83}, 054620
(2011).

\bibitem{bao} Z. H. Liu and J. D. Bao, Phys. Rev. C \textbf{84}, 031602(R)
(2011).

\bibitem{Gup05} Raj K. Gupta, Monika Manhas, et al., Phys. Rev. C \textbf{72}, 014607 (2005).

\bibitem{Zag08} V. Zagrebaev and W. Greiner, Phys. Rev. C \textbf{78}, 034610
(2008).

\bibitem{Ada09} G. G. Adamian, N. V. Antonenko, and V. V.
Sargsyan, Phys. Rev. C \textbf{79}, 054608 (2009).

\bibitem{Siw07} K. Siwek-Wilczy${\rm \acute{n}}$ska, I. Skwira-Chalot and J. Wilczy${\rm \acute{n}}$ski,
Int. J. Mod. Phys. E \textbf{16}, 483 (2007).

\bibitem{Nas} A. K. Nasirov, et al., Phys. Rev. C \textbf{84}, 044612
(2011).

\bibitem{Huang} M. Huang, Z. Gan, X. Zhou, J. Li, and W. Scheid, Phys. Rev. C \textbf{82}, 044614
(2010).

\bibitem{Zhou11}  B. N. Lu, E. G. Zhao, S. G. Zhou,
arXiv:nucl-th/1110.6769v1.


\bibitem{Dull} Ch. E. D\"ullmann, ``News from TASCA", talk given at the 10th Workshop on Recoil Separator for Superheavy
Element Chemistry, 2011, GSI.

\bibitem{Wang11} M. Liu, N. Wang, Y. Deng and X. Wu, Phys. Rev. C
\textbf{84}, 014333 (2011).

\bibitem{Love}W. Loveland, Phys. Rev. C \textbf{76}, 014612
(2007).

\bibitem{Wang08} N. Wang, K. Zhao, W. Scheid, and X. Wu, Phys. Rev. C \textbf{77}, 014603 (2008); http://www.imqmd.com/wangning/hivap2.rar.
\bibitem{Liu06}  M. Liu, N. Wang, Z. Li, X. Wu and E. Zhao, Nucl. Phys. A \textbf{768}, 80 (2006).
\bibitem{Wang06} N. Wang, X. Wu, Z. Li, M. Liu, and W. Scheid, Phys. Rev. C \textbf{74}, 044604
(2006).
\bibitem{Wang09} N. Wang, M. Liu and Y. Yang, Sci. China G \textbf{52}, 1554 (2009).



\bibitem{Reis81} W. Reisdorf, Z. Phys. A \textbf{300}, 227 (1981).
\bibitem{Reis85} W. Reisdorf, F. P. Hessberger, et al., Nucl. Phys. A \textbf{444}, 154 (1985).
\bibitem{Reis92} W. Reisdorf and M. Sch\"adel, Z. Phys. A \textbf{343}, 47 (1992).

\bibitem{Ogantalk} Y. Oganessian, ``SHE in JINR", talk given at the 109th Session of the JINR
Scientific Council, 2010, Dubna.

\bibitem{Pop} A. Popeko, ``Plans of Superheavy Elements Investigations in FLNR",
talk given at the Annual NuSTAR Meeting, 2010, GSI.

\bibitem{Bart82} J. Bartel, Ph. Quentin, M. Brack, C. Guet and H.B.
Hakansson, Nucl. Phys. A \textbf{386}, 79 (1982).

\bibitem{Moll95} P. M\"oller, J. R. Nix, W. D. Myers, W. J. Swiatecki, At. Data and
Nucl. Data Tables \textbf{59}, 185 (1995).

\bibitem{Tian07} J. Tian, N. Wang, Z. Li, Chin. Phys. Lett. \textbf{24}, 905 (2007), and references therein.

\bibitem{Myer67} W. D. Myers and W. J. Swiatecki, Ark. Fys. \textbf{36}, 342 (1967).

\bibitem{WN} Nan Wang, private communication.

\bibitem{Mosel} U. Mosel and W. Greiner, Z. Physik \textbf{222}, 261 (1969).

\bibitem{Smol} R. Smola${\rm \acute{n}}$czuk, J. Skalski, and A. Sobiczewski, Phys. Rev.
C \textbf{52}, 1871 (1995).

\end{thebibliography}
\end{document}